%% file: cuauv1.tex
\documentclass[aps,prl,twocolumn,superscriptaddress,nofootinbib,showpacs,showkeys,dvipdfmx,longbibliography]{revtex4-1}

\usepackage{graphicx}
\usepackage[mathlines]{lineno}
\usepackage{xspace}
\usepackage{url}
\usepackage{mediabb}
\usepackage{epstopdf}
\usepackage[normalem]{ulem}

\usepackage{color}
\definecolor{dgreen}{cmyk}{1.,0.,1.,0.2}        
\definecolor{orange}{cmyk}{0.,0.353,1.,0.}    

\def \snn{\mbox{$\sqrt{s_{_{NN}}}$}\xspace}

\newcommand{ \be }{\begin{eqnarray}}
\newcommand{ \ee }{\end{eqnarray}}

\newcommand{\pt}{\mbox{$p_T$}\xspace}
\newcommand{\mt}{\mbox{$m_T$}\xspace}

\begin{document}
\nolinenumbers

\preprint{\sl version 11, \today. }

\title{Charge-dependent directed flow in Cu+Au collisions at \snn = 200 GeV}

\input{authorlist_08122016_add.tex}


\date{\today}

\begin{abstract}
We present the first measurement of charge-dependent
directed flow in Cu+Au collisions at \snn = 200 GeV.  The results are
presented as a function of the particle transverse momentum and
pseudorapidity for different centralities. A finite difference between
the directed flow of positive and negative charged particles is
observed that qualitatively agrees with the expectations from the
effects of the initial strong electric field between two colliding
ions with different nuclear charges.  The measured difference in
directed flow is much smaller than that obtained from the
parton-hadron-string-dynamics (PHSD) model, 
which suggests that most of the electric charges, 
i.e. quarks and antiquarks, have not yet been created during
the lifetime of the strong electric field, which is of the order of, 
or less than, 1fm/$c$.
\end{abstract}

\maketitle

%
Hot and dense nuclear matter has been extensively studied in
nucleus-nucleus collisions at the Relativistic Heavy Ion Collider
(RHIC)~\cite{wh_phenix,wh_star,wh_phobos,wh_brahms} and the Large
Hadron Collider (LHC)~\cite{Raa_alice,jet_cms,dijet_atlas}.  Numerous
experimental results have suggested that a Quark-Gluon Plasma (QGP)
consisting of deconfined quarks and gluons is created in these
collisions.  At present, the emphasis is on characterizing the
detailed properties of the QGP.

One of the most important and informative experimental observables
used to study the properties of the QGP is the azimuthal
anisotropic flow which can be characterized by the Fourier
coefficients extracted from the azimuthal distribution of
the final state particles~\cite{Voloshin:2008dg}.
The second-order Fourier coefficient (so called
elliptic flow) and higher-order Fourier coefficients $v_{n}$ ($n>2$)
are found to be very sensitive to the shear viscosity over entropy
density ratio ($\eta$/s)~\cite{vnphenix,schenke}. The first-order
Fourier coefficient $v_{1}$, also known as directed flow, is sensitive
to the equation of state of the medium and therefore could be a
possible probe of a QGP phase
transition~\cite{v1_wiggle,v1_eos,BES_pv1}.

\begin{figure}[hbt]
\begin{center}
\includegraphics[width=0.93\linewidth,clip]{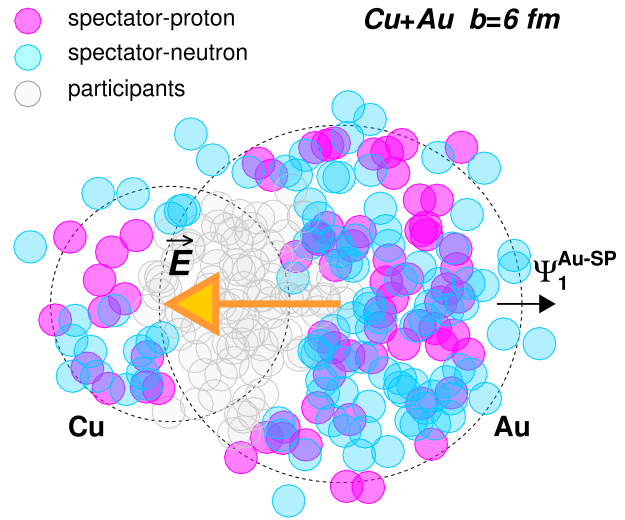}
\caption{\label{fig:cuau}(Color online) Example of a non-central Cu+Au
  collision viewed in the transverse plane showing an initial
  electric field $\vec{E}$ caused by the charge difference between two
  nuclei. $\Psi_1^{\rm Au\mathchar`-SP}$ denotes the direction of Au
  spectators.}
\end{center}
\end{figure}

Recent theoretical studies suggest that an asymmetric colliding system
can provide new insights regarding the properties of a QGP, 
such as the electric conductivity~\cite{hirono}
and the time evolution of the quark densities~\cite{voronyuk}.
Figure \ref{fig:cuau} shows an example of the distribution of
spectators and participants (protons and neutrons) in the transverse
plane for a Cu+Au collision assuming an impact parameter of 6 fm. Due
to the difference in the number of protons in the two nuclei, a strong
electric field is created at the initial stage of the collision and
the direction of the field is indicated by the arrow in
Fig.~\ref{fig:cuau}.
The lifetime of the field might be very short, of the order of a
fraction of a fm/$c$ (e.g. t$\sim$0.25 fm/$c$ from
Ref.~\cite{hirono,voronyuk}), but the electric charges from quarks and
antiquarks that are present in the early stage of the collision would
experience the Coulomb force and so would be pushed along or opposite to the
field direction depending on the particle charge.  The azimuthal distribution
of produced particles (including the effect of the electric field) can
be written as~\cite{deng,hirono}
\begin{equation}
\frac{dN^{\pm}}{d\phi} \propto 1+2v_{1}\cos(\phi-\Psi_{1}) \pm
2d_{E}\cos(\phi-\psi_{E}) \cdots,
\end{equation}
where $\phi$ is the azimuthal angle for a particle, $\Psi_{1}$ is the
angle of orientation for the first-order event plane, and the upper
(lower) sign of $\pm$ is for the positively (negatively) charged
particles. $\psi_{E}$ denotes the azimuthal angle of the electric
field; it is strongly correlated with $\Psi_1$ (see Figure 1)  
but can differ from $\Psi_1$ event-by-event due to the fluctuation 
of the initial nucleon distribution.
The coefficient $d_{E}$ characterizes the strength of dipole
deformation induced by the electric field and is proportional to the
electric conductivity of the plasma.  Then the directed flow, $v_{1}$,
of positively and negatively charged particles can be expressed as:
\begin{equation}
v_{1}^{\pm} = v_{1} \pm d_{E}\langle \cos(\Psi_{1}-\psi_{E}) \rangle,
\label{eq:v1ch}
\end{equation}
where $\langle\,\rangle$ means an average over all particles in all
events. Equation~(\ref{eq:v1ch}) illustrates how the presence of an
electric field results in charge separation for directed flow. The
strength of the charge separation depends on the number of
(anti)quarks existing at the earliest stages of the collision when the
electric field is strong. Therefore, the measurement of
charge-dependent directed flow can be used to test the quark
production mechanism, such as the two-wave scenario of quark
production~\cite{2wave1,2wave2}.
%
Also, understanding the time evolution of the quark density in
heavy-ion collisions is very important for a detailed theoretical
prediction of the Chiral Magnetic Effect~\cite{CME1,CME2} and the
Chiral Magnetic Wave~\cite{CMW_org,CMW2}. These effects are supposed
to emerge under an initial strong magnetic field and are actively
searched for by various
experiments~\cite{Kharzeev:2015znc,CME_star,CME_alice,CMW_star,CMW_alice}.

In this paper, we present the first measurement of the
charge-dependent directed flow in Cu+Au collisions at \snn = 200 GeV.
The results are presented for different collision centralities as a
function of the particle transverse momentum (\pt) and pseudorapidity ($\eta$).
For comparison we also show results for Au+Au collisions where
the effect is expected to be significantly smaller, because the average
electric field in these collisions is expected to be zero 

The data reported in this analysis are from Cu+Au collisions at \snn =
200 GeV collected in 2012 with the STAR detector.
The collision vertices were reconstructed using
charged-particle tracks measured in the Time Projection Chamber
(TPC)~\cite{tpc}. The TPC covers the full azimuth and has a pseudorapidity
range of $|\eta|<1.0$.  Events were selected to have the collision
vertex position within $\pm$ 30~cm from the center of the TPC in the beam
direction and within $\pm$ 2~cm in the radial direction with respect to the
center of the beam.  An additional constraint on the vertex position
along the beam direction was imposed using the Vertex Position
Detector (VPD)~\cite{vpd} to reduce the beam-induced
background. Forty-four million minimum bias Cu+Au events were used in
the analysis, where the minimum bias trigger required hits of VPDs and 
Zero Degree Calorimeters (described below) in Cu and Au going directions.  
In addition, ninety-five million minimum bias Au+Au
events, collected in 2010, were analyzed in the same way for
comparison.

\begin{figure*}[ht]
\begin{center}
\includegraphics[width=0.92\textwidth]{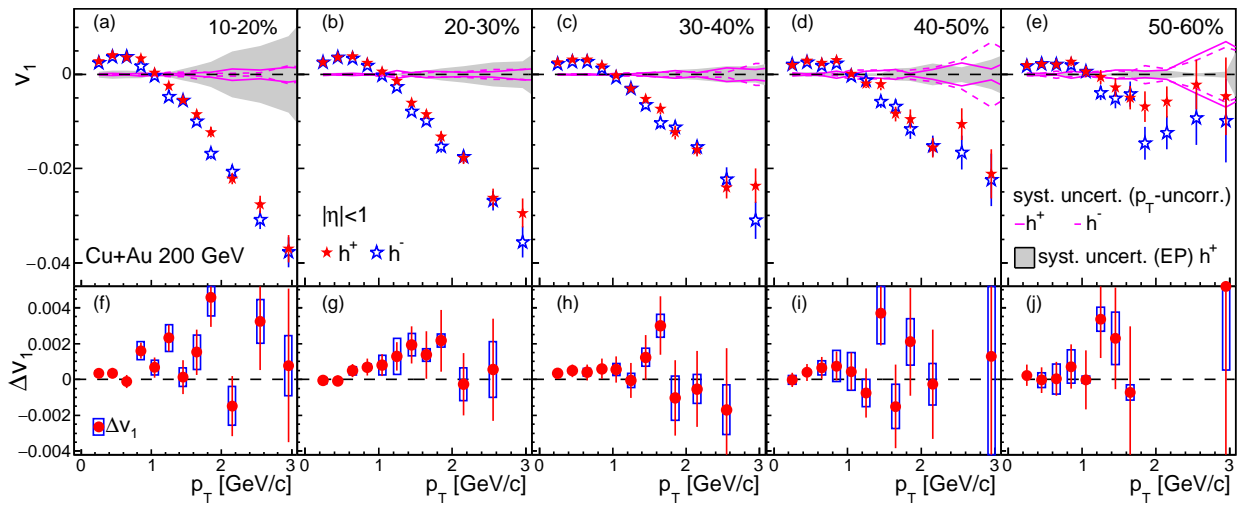}
\caption{\label{fig:v1pt_10step}(Color online) Directed flow of
  positive and negative particles from minimum bias Cu+Au collisions
  at \snn = 200 GeV, as a function of \pt, in five centrality bins.  The
  difference between the positive and negative spectra is shown in the
  lower panels, where the open boxes show the systematic
  uncertainties.  See text for the definition of the positive
  direction for $v_1$.}
\end{center}
\end{figure*}

The centrality of each collision was determined by measuring
event-by-event multiplicity and interpreting the measurement with a
tuned Monte Carlo Glauber calculation~\cite{glauber,BESv2}.  The
first-order event plane was determined by Zero Degree Calorimeters
(ZDC) that are equipped with shower maximum detectors
(SMD)~\cite{zdc,v1smd,v1smd_star}.  The ZDC-SMDs are located at
forward and backward angles ($|\eta|>6.3$) and they measure the energy
deposited by spectator neutrons as well as the transverse
distribution of the neutrons.  It is worth noting that spectator neutrons, on average,
deflect outward from the center-line of the collisions~\cite{spflow}
and thus provide information on the direction of the electric field.  The
event plane resolution was estimated by the three-subevent
method~\cite{TwoSub}.  It reaches a maximum of 0.26 for mid-central
events when using the ZDC-SMD in the Au-going direction.  Analyzed
tracks were required to have the
distance of closest approach to the primary vertex to be less than
3~cm, and have at least 15 TPC space points used in their
reconstruction. Furthermore, the ratio of the number of fit points to
maximum possible number of TPC space points (45) was required to be larger
than~0.52 to remove split tracks.  The \pt of tracks was limited to the range
$0.15<\pt<5$~GeV/$c$.

Directed flow, $v_{1}$, was measured using the ``event plane method": 
\begin{equation}
v_{1} = \langle\cos (\phi-\Psi_{1})\rangle/{\rm Res}\{\Psi_{1}\},
\end{equation}
where $\phi$ is the azimuthal angle of a track and 
${\rm Res}\{\Psi_{1}\}$ denotes the event plane resolution.  
Directed flow is measured with respect to the spectator plane determined by
  the ZDC-SMD in the Au-going direction, but the sign of $\Psi_{1}$ is
  defined to be positive at forward rapidities (Cu-going direction) to
  keep the convention of past $v_{1}$ measurements. Note that
  $v_{1}(\eta)$, measured with respect to the spectator plane of one of
  the nuclei, includes the component due to density
  fluctuations~\cite{Teaney,Luzum,aliceV1} and 
  does not necessarily
cross zero at $\eta=0$ even for symmetric collisions.  
Also, note that $\eta$ is measured in the nucleon-nucleon center-of-mass frame.

Systematic uncertainties in the results have been estimated by
variation of the size of the collision $z$-vertex window, variation of
the track quality cuts, and by using different combinations of the
three-subevents in the estimation of the event plane resolution. The
relative systematic uncertainties associated with the z-vertex and
track quality cuts are below 6\% for mid-central events and were found
to be uncorrelated in \pt. The uncertainty of the event plane
resolution was studied by varying the detector combinations used in
the three-subevent method. The detector choices were two ZDC-SMD
detectors, and one of two Beam Beam Counters~\cite{bbc} located at
forward and backward angles ($3.3<|\eta|<5$), or the endcap
electromagnetic calorimeter ($1.086<\eta<2$)~\cite{eemc}.  The
associated systematic uncertainty is \pt-correlated; namely, all data
points move in the same direction as the sign of $v_1$ with the same
fraction. The change in $v_{1}$ due to the use of different subevents
is $\sim$7\% for mid-central events and increases up to 22\% for more
central and peripheral events.  This is the largest systematic
uncertainty in these measurements. 
The Cu+Au data were taken only with one polarity of the magnetic field. 
In order to check the effect of the magnetic field, the Au+Au data were also analyzed 
with the same polarity, where no effect has been observed.

Figure~\ref{fig:v1pt_10step}(a-e) shows the directed flow of positive
($h^{+}$) and negative ($h^{-}$) charged particles as a function of
\pt for five different centrality bins. The solid (dashed) lines
around $v_{1}=0$ show the \pt-uncorrelated systematic uncertainties
and the shaded bands, indicated with ``EP", show the \pt-correlated
systematic uncertainties associated with the event plane resolution.
  The observed $v_{1}$ has positive value at low \pt
($\pt<1$ GeV/$c$) and goes negative at high \pt. 

\begin{figure}[t]
\begin{center}
\includegraphics[width=0.95\linewidth]{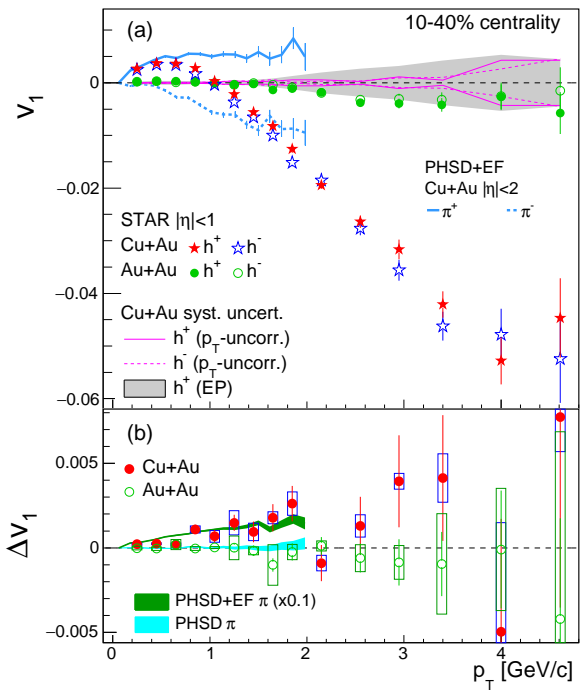}
\caption{\label{fig:v1pt_10-40}(Color online) Directed flow of
  positive and negative particles and the difference between the two
  spectra as a function of \pt in 10\%-40\% centrality in Cu+Au and
  Au+Au collisions.  The PHSD model calculations~\cite{voronyuk} for
  charged pions with and without the initial electric field (EF) in
  the same centrality region are presented for comparison. Note that the charge
  difference of $v_{1}$ with the EF-on is scaled by 0.1.}
\end{center}
\end{figure}

The trend of the $p_T$ dependence is similar to that of $v_1$ measured 
in symmetric collisions~\cite{aliceV1,v1smd_star}.
The magnitude of our $v_1$ is about ten times larger than $v_1^{\rm even}$ 
(as shown in Fig.~\ref{fig:v1pt_10-40}) and twice (ten times) larger than $v_1^{\rm odd}$ 
at \pt = 1 (3) GeV/$c$ in Au+Au collisions~\cite{v1smd_star}.
This is likely because the $v_{1}$ in symmetric collisions originates only 
from the density fluctuations, while the $v_{1}$ in asymmetric collisions is
dominated by the initial density asymmetry~\cite{bozek,v1_heinz}. The
average of $v_1$ for positive and negative particles is consistent,
within errors, with the results of charge-combined directed flow
measurements recently published by the PHENIX
Collaboration~\cite{cuauphenix}.

Figure~\ref{fig:v1pt_10step}(f-j) shows $\Delta v_{1}$ defined as the
difference in $v_{1}$ between positive and negative charged particles.
Note that the large uncertainties on the event plane resolution largely cancel
out in $\Delta v_1$.  The difference, $\Delta v_{1}$, is about 10\% of
$v_{1}$ in magnitude. It tends to be positive for $\pt<2$~GeV/$c$ in
10\%-30\% centrality and becomes consistent with zero by 50\%-60\%
centrality within large systematic uncertainties.  The small but
finite $\Delta v_{1}$ agrees with the expectation for the effects of
the initial electric field.  The sign flipping of the electric
  field discussed in Ref.~\cite{hirono} seems not to be observed within the current
  uncertainty, which is close to the expectation discussed in Ref.~\cite{deng}.

Figure~\ref{fig:v1pt_10-40} shows $v_{1}$ and $\Delta v_{1}$ in the
10\%-40\% centrality bin. For $\pt<2$~GeV/$c$, the $\Delta v_{1}$
seems to increase with \pt.  The $v_{1}$ results from Au+Au collisions (the
so-called even component of $v_1$) show much smaller 
values ($\sim$by a factor of 10) compared to those in Cu+Au. 
Note that the odd component of $v_1$ in Au+Au collisions
  is similarly small~\cite{v1smd_star}.  The $\Delta v_1$ in Au+Au is
consistent with zero.  Calculations for charged pions from the
parton-hadron-string-dynamics (PHSD) model~\cite{voronyuk}, which is a
dynamical transport approach in the partonic and hadronic phases, are
compared to the data. As indicated in Eq.~\ref{eq:v1ch}, the measured 
$\Delta v_1$ could be smeared by the fluctuations in $\psi_{E}$ and 
$\Psi_1$ orientations, but note that the PHSD model takes  
such event-by-event fluctuations into account.
The PHSD model calculates two cases:
charge-dependent $v_{1}$ with and without the initial electric field
(EF). For the case with the EF switched on, the model assumes that all
electric charges are affected by the EF and this results in a large
separation of $v_{1}$ between positive and negative particles as shown
in Fig.~\ref{fig:v1pt_10-40}(a).  In Fig.~\ref{fig:v1pt_10-40}(b), the
calculations of the $\Delta v_{1}$ with and without the EF are shown
together, but note that the EF-on data points are scaled by 0.1
relative to the PHSD results. After scaling by 0.1, the model
describes rather well the \pt dependence of the measured data for
$\pt<2$~GeV/$c$.

The magnitude of $\Delta v_{1}$ should depend on the number of quarks
and antiquarks and the electric conductivity at the time when the EF is
strong.
We note, however, that the electric conductivity calculations in lattice QCD
differ by an order of magnitude between different
groups~\cite{Gupta,Aarts}, and the perturbative QCD
calculations~\cite{pQCD,pQCD2} predict larger values
than lattice QCD. In comparison, the electric conductivity evaluated in the PHSD
model is close to the lower value of the lattice QCD
calculations~\cite{phsd_Sgm}.  Therefore, the fact that the observed $\Delta
v_{1}$ is ten times smaller compared to the PHSD model calculation with EF-on likely indicates 
a small number of quarks and antiquarks at $t\leq0.25$~fm/$c$.
The lifetime of the electric field could be longer 
if the created medium is a good conductor~\cite{McLerrana2014,KTuchin2013,BZakharov2014}. 
Therefore the fraction of quarks present at the early times could be even lower.

\begin{figure}[t]
\begin{center}
\includegraphics[width=\linewidth]{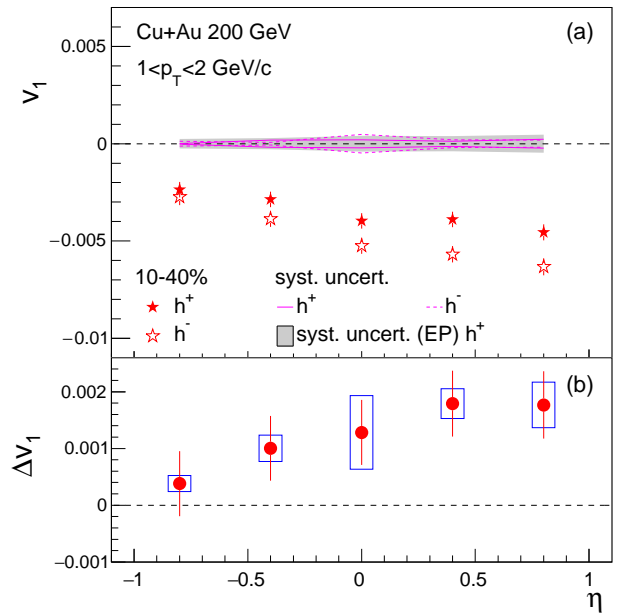}
\caption{\label{fig:v1eta_10-40}(Color online) Directed flow of
  positive and negative particles and the difference between the two
  spectra as a function of $\eta$ in the 10\%-40\% centrality bin, where
  positive $\eta$ denotes the Cu-going direction. }
\end{center}
\end{figure}
We can roughly estimate the ratio of the number of (anti)quarks that
existed at very early times, to the total number of (anti)quarks
created in the collision (final state number) using the parton
distribution functions (PDF).  We have used the HERAPDF1.5
(NLO~$Q^2=4$~ GeV$^{2}$) parton distribution functions~\cite{herapdf},
and we assumed that the number of quarks in the initial state
corresponds to the number of quarks given by the PDF in the
corresponding momentum region.  We assume that the total number
  of hadrons in the final state is approximately equal to the number of
  partons in the initial state; meaning that one gluon in the
  initial state corresponds approximately to two quarks in the final
  state.  
  Then, the ratio of the initial quarks to
the total number of quarks created in the collision can be calculated
by comparing the PDFs at $x \sim \mt e^{\eta}/\sqrt{s}\approx 0.01$ 
corresponding to the kinematics of the current measurement.
Note that such an estimate depends very weakly on the exact momentum
fraction range $x$ and $Q^2$.
Using this approach we find that the ratio is about 0.15, 
close to the scale factor applied to the
PHSD model calculations shown in Fig.~\ref{fig:v1pt_10-40}.

The pseudorapidity dependence of $v_{1}$ and $\Delta v_{1}$ was
measured in the 10\%-40\% centrality bin as shown in
Fig.~\ref{fig:v1eta_10-40}.  As seen in Fig.~\ref{fig:v1pt_10step} and
Fig.~\ref{fig:v1pt_10-40}, the $\Delta v_{1}$ exhibits a stronger
signal in the $1<\pt<2$~GeV/$c$ range. Therefore in
Fig.~\ref{fig:v1eta_10-40} the signal is integrated over that
range. The magnitude of $v_{1}$ becomes larger at forward rapidities
and $\Delta v_{1}$ has a finite value within $|\eta|<1$.  The
difference, $\Delta v_{1}$, seems to be larger at forward rapidities
(Cu-going direction). This might be related to the
$v_1(y)$ (where $y$ denotes rapidity) slope difference between
particles and antiparticles~\cite{BES_pv1} and the shift of the
center-of-mass in asymmetric collisions, although the uncertainty of
the data is still too large to discuss this in detail.

In conclusion, we have presented results for the first measurements of
charge-dependent directed flow in Cu+Au collisions at \snn =
200~GeV.  A finite difference in $v_1$ between positive and negative
charged particles was observed in the transverse momentum range of
$0.15<\pt<2$~GeV/$c$ and the pseudorapidity range of $|\eta|<1$. These
results are consistent with the presumption 
of a strong, initial, electric field in asymmetric collisions. 
The observed $\Delta v_{1}$ was compared to the PHSD
model calculations that includes the effect of an electric field.
The \pt dependence of $\Delta v_{1}$ is qualitatively described by the
model in the region less than 2 GeV/$c$.  However the magnitude of
$\Delta v_{1}$ is smaller by a factor of 10 than the model
predictions, assuming that all quarks are created at the initial
time. This may indicate that most of quarks and antiquarks have not
yet been created within the lifetime of the electric field
($t\leq0.25$~fm/$c$).
A simple estimate of the fraction of the initial quarks present in the
participant nucleons, relative to all quarks created during the
collision (assuming each gluon to be converted to a $q\bar{q}$-pair),
is consistent with this interpretation.  These results provide
important information for understanding the time evolution of particle
production, and will constrain estimates of the magnitude of the
Chiral Magnetic Effect and Chiral Magnetic Wave induced by the initial
strong magnetic field.

\begin{acknowledgements}
We thank the RHIC Operations Group and RCF at BNL, the NERSC Center at
LBNL, the KISTI Center in Korea, and the Open Science Grid consortium
for providing resources and support. This work was supported in part
by the Office of Nuclear Physics within the U.S. DOE Office of
Science, the U.S. NSF, the Ministry of Education and Science of the
Russian Federation, NSFC, CAS, MoST and MoE of China, the National
Research Foundation of Korea, NCKU (Taiwan), GA and MSMT of the Czech
Republic, FIAS of Germany, DAE, DST, and UGC of India, the National
Science Centre of Poland, National Research Foundation, the Ministry
of Science, Education and Sports of the Republic of Croatia, and
RosAtom of Russia.
\end{acknowledgements}

\bibliography{ref_cuauv1}

\end{document}

%% file: authorlist_08122016_add.tex
\affiliation{AGH University of Science and Technology, FPACS, Cracow 30-059, Poland}
\affiliation{Argonne National Laboratory, Argonne, Illinois 60439}
\affiliation{Brookhaven National Laboratory, Upton, New York 11973}
\affiliation{University of California, Berkeley, California 94720}
\affiliation{University of California, Davis, California 95616}
\affiliation{University of California, Los Angeles, California 90095}
\affiliation{Central China Normal University, Wuhan, Hubei 430079}
\affiliation{University of Illinois at Chicago, Chicago, Illinois 60607}
\affiliation{Creighton University, Omaha, Nebraska 68178}
\affiliation{Czech Technical University in Prague, FNSPE, Prague, 115 19, Czech Republic}
\affiliation{Nuclear Physics Institute AS CR, 250 68 Prague, Czech Republic}
\affiliation{Frankfurt Institute for Advanced Studies FIAS, Frankfurt 60438, Germany}
\affiliation{Institute of Physics, Bhubaneswar 751005, India}
\affiliation{Indian Institute of Technology, Mumbai 400076, India}
\affiliation{Indiana University, Bloomington, Indiana 47408}
\affiliation{Alikhanov Institute for Theoretical and Experimental Physics, Moscow 117218, Russia}
\affiliation{University of Jammu, Jammu 180001, India}
\affiliation{Joint Institute for Nuclear Research, Dubna, 141 980, Russia}
\affiliation{Kent State University, Kent, Ohio 44242}
\affiliation{University of Kentucky, Lexington, Kentucky, 40506-0055}
\affiliation{Lamar University, Physics Department, Beaumont, Texas 77710}
\affiliation{Institute of Modern Physics, Chinese Academy of Sciences, Lanzhou, Gansu 730000}
\affiliation{Lawrence Berkeley National Laboratory, Berkeley, California 94720}
\affiliation{Lehigh University, Bethlehem, PA, 18015}
\affiliation{Max-Planck-Institut fur Physik, Munich 80805, Germany}
\affiliation{Michigan State University, East Lansing, Michigan 48824}
\affiliation{National Research Nuclear University MEPhI, Moscow 115409, Russia}
\affiliation{National Institute of Science Education and Research, Bhubaneswar 751005, India}
\affiliation{National Cheng Kung University, Tainan 70101 }
\affiliation{Ohio State University, Columbus, Ohio 43210}
\affiliation{Institute of Nuclear Physics PAN, Cracow 31-342, Poland}
\affiliation{Panjab University, Chandigarh 160014, India}
\affiliation{Pennsylvania State University, University Park, Pennsylvania 16802}
\affiliation{Institute of High Energy Physics, Protvino 142281, Russia}
\affiliation{Purdue University, West Lafayette, Indiana 47907}
\affiliation{Pusan National University, Pusan 46241, Korea}
\affiliation{Rice University, Houston, Texas 77251}
\affiliation{University of Science and Technology of China, Hefei, Anhui 230026}
\affiliation{Shandong University, Jinan, Shandong 250100}
\affiliation{Shanghai Institute of Applied Physics, Chinese Academy of Sciences, Shanghai 201800}
\affiliation{State University Of New York, Stony Brook, NY 11794}
\affiliation{Temple University, Philadelphia, Pennsylvania 19122}
\affiliation{Texas A\&M University, College Station, Texas 77843}
\affiliation{University of Texas, Austin, Texas 78712}
\affiliation{University of Houston, Houston, Texas 77204}
\affiliation{Tsinghua University, Beijing 100084}
\affiliation{Southern Connecticut State University, New Haven, CT, 06515}
\affiliation{United States Naval Academy, Annapolis, Maryland, 21402}
\affiliation{Valparaiso University, Valparaiso, Indiana 46383}
\affiliation{Variable Energy Cyclotron Centre, Kolkata 700064, India}
\affiliation{Warsaw University of Technology, Warsaw 00-661, Poland}
\affiliation{Wayne State University, Detroit, Michigan 48201}
\affiliation{World Laboratory for Cosmology and Particle Physics (WLCAPP), Cairo 11571, Egypt}
\affiliation{Yale University, New Haven, Connecticut 06520}

\author{L.~Adamczyk}\affiliation{AGH University of Science and Technology, FPACS, Cracow 30-059, Poland}
\author{J.~K.~Adkins}\affiliation{University of Kentucky, Lexington, Kentucky, 40506-0055}
\author{G.~Agakishiev}\affiliation{Joint Institute for Nuclear Research, Dubna, 141 980, Russia}
\author{M.~M.~Aggarwal}\affiliation{Panjab University, Chandigarh 160014, India}
\author{Z.~Ahammed}\affiliation{Variable Energy Cyclotron Centre, Kolkata 700064, India}
\author{I.~Alekseev}\affiliation{Alikhanov Institute for Theoretical and Experimental Physics, Moscow 117218, Russia}\affiliation{National Research Nuclear University MEPhI, Moscow 115409, Russia}
\author{D.~M.~Anderson}\affiliation{Texas A\&M University, College Station, Texas 77843}
\author{R.~Aoyama}\affiliation{Brookhaven National Laboratory, Upton, New York 11973}
\author{A.~Aparin}\affiliation{Joint Institute for Nuclear Research, Dubna, 141 980, Russia}
\author{D.~Arkhipkin}\affiliation{Brookhaven National Laboratory, Upton, New York 11973}
% ...

\collaboration{STAR Collaboration}\noaffiliation

%% file: cuauv1.bbl
\begin{thebibliography}{53}%
\makeatletter
\providecommand \@ifxundefined [1]{%
 \@ifx{#1\undefined}
}%
\providecommand \@ifnum [1]{%
 \ifnum #1\expandafter \@firstoftwo
 \else \expandafter \@secondoftwo
 \fi
}%
\providecommand \@ifx [1]{%
 \ifx #1\expandafter \@firstoftwo
 \else \expandafter \@secondoftwo
 \fi
}%
\providecommand \natexlab [1]{#1}%
\providecommand \enquote  [1]{``#1''}%
\providecommand \bibnamefont  [1]{#1}%
\providecommand \bibfnamefont [1]{#1}%
\providecommand \citenamefont [1]{#1}%
\providecommand \href@noop [0]{\@secondoftwo}%
\providecommand \href [0]{\begingroup \@sanitize@url \@href}%
\providecommand \@href[1]{\@@startlink{#1}\@@href}%
\providecommand \@@href[1]{\endgroup#1\@@endlink}%
\providecommand \@sanitize@url [0]{\catcode `\\12\catcode `\$12\catcode
  `\&12\catcode `\#12\catcode `\^12\catcode `\_12\catcode `\%12\relax}%
\providecommand \@@startlink[1]{}%
\providecommand \@@endlink[0]{}%
\providecommand \url  [0]{\begingroup\@sanitize@url \@url }%
\providecommand \@url [1]{\endgroup\@href {#1}{\urlprefix }}%
\providecommand \urlprefix  [0]{URL }%
\providecommand \Eprint [0]{\href }%
\providecommand \doibase [0]{http://dx.doi.org/}%
\providecommand \selectlanguage [0]{\@gobble}%
\providecommand \bibinfo  [0]{\@secondoftwo}%
\providecommand \bibfield  [0]{\@secondoftwo}%
\providecommand \translation [1]{[#1]}%
\providecommand \BibitemOpen [0]{}%
\providecommand \bibitemStop [0]{}%
\providecommand \bibitemNoStop [0]{.\EOS\space}%
\providecommand \EOS [0]{\spacefactor3000\relax}%
\providecommand \BibitemShut  [1]{\csname bibitem#1\endcsname}%
\let\auto@bib@innerbib\@empty
\bibitem [{\citenamefont {{K. Adcox {\it et al.} (PHENIX
  Collaboration)}}(2005)}]{wh_phenix}%
  \BibitemOpen
  \bibfield  {author} {\bibinfo {author} {\bibnamefont {{K. Adcox {\it et al.}
  (PHENIX Collaboration)}}},\ }\bibfield  {title} {\enquote {\bibinfo {title}
  {{Formation of dense partonic matter in relativistic nucleus-nucleus
  collisions at RHIC: Experimental evaluation by the PHENIX collaboration}},}\
  }\href@noop {} {\bibfield  {journal} {\bibinfo  {journal} {Nucl. Phys.}\
  }\textbf {\bibinfo {volume} {A757}},\ \bibinfo {pages} {184} (\bibinfo {year}
  {2005})}\BibitemShut {NoStop}%
\bibitem [{\citenamefont {{J. Adams {\it et al.} (STAR
  Collaboration)}}(2005)}]{wh_star}%
  \BibitemOpen
  \bibfield  {author} {\bibinfo {author} {\bibnamefont {{J. Adams {\it et al.}
  (STAR Collaboration)}}},\ }\bibfield  {title} {\enquote {\bibinfo {title}
  {{Experimental and Theoretical Challenges in the Search for the Quark Gluon
  Plasma: The STAR Collaboration's Critical Assessment of the Evidence from
  RHIC Collisions}},}\ }\href@noop {} {\bibfield  {journal} {\bibinfo
  {journal} {Nucl. Phys.}\ }\textbf {\bibinfo {volume} {A757}},\ \bibinfo
  {pages} {102} (\bibinfo {year} {2005})}\BibitemShut {NoStop}%
\bibitem [{\citenamefont {{B. B. Back {\it et al.} (PHOBOS
  Collaboration)}}(2005)}]{wh_phobos}%
  \BibitemOpen
  \bibfield  {author} {\bibinfo {author} {\bibnamefont {{B. B. Back {\it et
  al.} (PHOBOS Collaboration)}}},\ }\bibfield  {title} {\enquote {\bibinfo
  {title} {{The PHOBOS Perspective on Discoveries at RHIC}},}\ }\href@noop {}
  {\bibfield  {journal} {\bibinfo  {journal} {Nucl. Phys.}\ }\textbf {\bibinfo
  {volume} {A757}},\ \bibinfo {pages} {28} (\bibinfo {year}
  {2005})}\BibitemShut {NoStop}%
\bibitem [{\citenamefont {{l. Arsene {\it et al.} (BRAHMS
  Collaboration)}}(2005)}]{wh_brahms}%
  \BibitemOpen
  \bibfield  {author} {\bibinfo {author} {\bibnamefont {{l. Arsene {\it et al.}
  (BRAHMS Collaboration)}}},\ }\bibfield  {title} {\enquote {\bibinfo {title}
  {{Quark Gluon Plasma an Color Glass Condensate at RHIC? The perspective from
  the BRAHMS experiment}},}\ }\href@noop {} {\bibfield  {journal} {\bibinfo
  {journal} {Nucl. Phys.}\ }\textbf {\bibinfo {volume} {A757}},\ \bibinfo
  {pages} {1} (\bibinfo {year} {2005})}\BibitemShut {NoStop}%
\bibitem [{\citenamefont {{K. Aamodt {\it et al.} (ALICE
  Collaboration)}}(2011)}]{Raa_alice}%
  \BibitemOpen
  \bibfield  {author} {\bibinfo {author} {\bibnamefont {{K. Aamodt {\it et al.}
  (ALICE Collaboration)}}},\ }\bibfield  {title} {\enquote {\bibinfo {title}
  {{Suppression of charged particle production at large transverse momentum in
  central Pb+Pb collisions at $\sqrt{s_{NN}}$ = 2.76 TeV}},}\ }\href@noop {}
  {\bibfield  {journal} {\bibinfo  {journal} {Phys. Lett.}\ }\textbf {\bibinfo
  {volume} {B696}},\ \bibinfo {pages} {30--39} (\bibinfo {year}
  {2011})}\BibitemShut {NoStop}%
\bibitem [{\citenamefont {{S. Chatrchyan {\it et al.} (CMS
  Collaboration)}}(2011)}]{jet_cms}%
  \BibitemOpen
  \bibfield  {author} {\bibinfo {author} {\bibnamefont {{S. Chatrchyan {\it et
  al.} (CMS Collaboration)}}},\ }\bibfield  {title} {\enquote {\bibinfo {title}
  {{Observation and studies of jet quenching in PbPb collisions at
  $\sqrt{s_{NN}}$ = 2.76 TeV}},}\ }\href@noop {} {\bibfield  {journal}
  {\bibinfo  {journal} {Phys. Rev.}\ }\textbf {\bibinfo {volume} {C84}},\
  \bibinfo {pages} {024906} (\bibinfo {year} {2011})}\BibitemShut {NoStop}%
\bibitem [{\citenamefont {{G. Aad {\it et al.} (ATLAS
  Collaboration)}}(2010)}]{dijet_atlas}%
  \BibitemOpen
  \bibfield  {author} {\bibinfo {author} {\bibnamefont {{G. Aad {\it et al.}
  (ATLAS Collaboration)}}},\ }\bibfield  {title} {\enquote {\bibinfo {title}
  {{Observation of a centrality-dependent dijet asymmetry in lead-lead
  collisions at $\sqrt{s_{NN}}$ = 2.76 TeV with the ATLAS detector}},}\
  }\href@noop {} {\bibfield  {journal} {\bibinfo  {journal} {Phys. Rev. Lett.}\
  }\textbf {\bibinfo {volume} {105}},\ \bibinfo {pages} {252303} (\bibinfo
  {year} {2010})}\BibitemShut {NoStop}%
\bibitem [{\citenamefont {Voloshin}\ \emph {et~al.}(2008)\citenamefont
  {Voloshin}, \citenamefont {Poskanzer},\ and\ \citenamefont
  {Snellings}}]{Voloshin:2008dg}%
  \BibitemOpen
  \bibfield  {author} {\bibinfo {author} {\bibfnamefont {S..~A}\ \bibnamefont
  {Voloshin}}, \bibinfo {author} {\bibfnamefont {A.~M}\ \bibnamefont
  {Poskanzer}}, \ and\ \bibinfo {author} {\bibfnamefont {R.}~\bibnamefont
  {Snellings}},\ }\bibfield  {title} {\enquote {\bibinfo {title} {{Collective
  phenomena in non-central nuclear collisions}},}\ }\href@noop {} {\  (\bibinfo
  {year} {2008})},\ \Eprint {http://arxiv.org/abs/0809.2949} {arXiv:0809.2949
  [nucl-ex]} \BibitemShut {NoStop}%
\bibitem [{\citenamefont {{A. Adare {\it et al.} (PHENIX
  Collaboration)}}(2011)}]{vnphenix}%
  \BibitemOpen
  \bibfield  {author} {\bibinfo {author} {\bibnamefont {{A. Adare {\it et al.}
  (PHENIX Collaboration)}}},\ }\bibfield  {title} {\enquote {\bibinfo {title}
  {{Measurements of Higher Order Flow Harmonics in Au+Au collisions at
  $\sqrt{s_{_{NN}}}$ = 200 GeV}},}\ }\href@noop {} {\bibfield  {journal}
  {\bibinfo  {journal} {Phys. Rev. Lett.}\ }\textbf {\bibinfo {volume} {107}},\
  \bibinfo {pages} {252301} (\bibinfo {year} {2011})}\BibitemShut {NoStop}%
\bibitem [{\citenamefont {Gale}\ \emph {et~al.}(2013)\citenamefont {Gale},
  \citenamefont {Jeon}, \citenamefont {Schenke}, \citenamefont {Tribedy},\ and\
  \citenamefont {Venugopalan}}]{schenke}%
  \BibitemOpen
  \bibfield  {author} {\bibinfo {author} {\bibfnamefont {C.}~\bibnamefont
  {Gale}}, \bibinfo {author} {\bibfnamefont {S.}~\bibnamefont {Jeon}}, \bibinfo
  {author} {\bibfnamefont {B.}~\bibnamefont {Schenke}}, \bibinfo {author}
  {\bibfnamefont {P.}~\bibnamefont {Tribedy}}, \ and\ \bibinfo {author}
  {\bibfnamefont {R.}~\bibnamefont {Venugopalan}},\ }\bibfield  {title}
  {\enquote {\bibinfo {title} {{Event-by-event anisotropic flow in heavy-ion
  collisions from combined Tang-Mills and viscous fluid dynamics}},}\
  }\href@noop {} {\bibfield  {journal} {\bibinfo  {journal} {Phys. Rev. Lett.}\
  }\textbf {\bibinfo {volume} {110}},\ \bibinfo {pages} {012302} (\bibinfo
  {year} {2013})}\BibitemShut {NoStop}%
\bibitem [{\citenamefont {Csernai}\ and\ \citenamefont {R$\ddot{\rm
  o}$hrich}(1999)}]{v1_wiggle}%
  \BibitemOpen
  \bibfield  {author} {\bibinfo {author} {\bibfnamefont {L.~P.}\ \bibnamefont
  {Csernai}}\ and\ \bibinfo {author} {\bibfnamefont {D.}~\bibnamefont
  {R$\ddot{\rm o}$hrich}},\ }\bibfield  {title} {\enquote {\bibinfo {title}
  {{Third flow component as QGP signal}},}\ }\href@noop {} {\bibfield
  {journal} {\bibinfo  {journal} {Phys. Lett.}\ }\textbf {\bibinfo {volume}
  {B458}},\ \bibinfo {pages} {454} (\bibinfo {year} {1999})}\BibitemShut
  {NoStop}%
\bibitem [{\citenamefont {{J. Brachmann and S. Soff and A. Dumitru and H.
  St$\ddot{\rm o}$cker and J. A. Maruhn and W. Greiner and L. V. Bravina and
  and D. H. Rischke}}(2000)}]{v1_eos}%
  \BibitemOpen
  \bibfield  {author} {\bibinfo {author} {\bibnamefont {{J. Brachmann and S.
  Soff and A. Dumitru and H. St$\ddot{\rm o}$cker and J. A. Maruhn and W.
  Greiner and L. V. Bravina and and D. H. Rischke}}},\ }\bibfield  {title}
  {\enquote {\bibinfo {title} {{Antiflow of nucleons at the softest point of
  the EoS}},}\ }\href@noop {} {\bibfield  {journal} {\bibinfo  {journal} {Phys.
  Rev.}\ }\textbf {\bibinfo {volume} {C61}},\ \bibinfo {pages} {024909}
  (\bibinfo {year} {2000})}\BibitemShut {NoStop}%
\bibitem [{\citenamefont {{L. Adamczyk {\it et al.} (STAR
  Collaboration)}}(2014)}]{BES_pv1}%
  \BibitemOpen
  \bibfield  {author} {\bibinfo {author} {\bibnamefont {{L. Adamczyk {\it et
  al.} (STAR Collaboration)}}},\ }\bibfield  {title} {\enquote {\bibinfo
  {title} {{Beam-Energy Dependence of the Directed Flow of Protons,
  Antiprotons, and Pions in Au+Au Collisions}},}\ }\href@noop {} {\bibfield
  {journal} {\bibinfo  {journal} {Phys. Rev. Lett.}\ }\textbf {\bibinfo
  {volume} {112}},\ \bibinfo {pages} {162301} (\bibinfo {year}
  {2014})}\BibitemShut {NoStop}%
\bibitem [{\citenamefont {Hirono}\ \emph {et~al.}(2014)\citenamefont {Hirono},
  \citenamefont {Hongo},\ and\ \citenamefont {Hirano}}]{hirono}%
  \BibitemOpen
  \bibfield  {author} {\bibinfo {author} {\bibfnamefont {Y.}~\bibnamefont
  {Hirono}}, \bibinfo {author} {\bibfnamefont {M.}~\bibnamefont {Hongo}}, \
  and\ \bibinfo {author} {\bibfnamefont {T.}~\bibnamefont {Hirano}},\
  }\bibfield  {title} {\enquote {\bibinfo {title} {{Estimation of the electric
  conductivity of the quark gluon plasma via asymmetric heavy-ion
  collisions}},}\ }\href@noop {} {\bibfield  {journal} {\bibinfo  {journal}
  {Phys. Rev.}\ }\textbf {\bibinfo {volume} {C90}},\ \bibinfo {pages} {021903}
  (\bibinfo {year} {2014})}\BibitemShut {NoStop}%
\bibitem [{\citenamefont {Voronyuk}\ \emph {et~al.}(2014)\citenamefont
  {Voronyuk}, \citenamefont {Toneev}, \citenamefont {Voloshin},\ and\
  \citenamefont {Cassing}}]{voronyuk}%
  \BibitemOpen
  \bibfield  {author} {\bibinfo {author} {\bibfnamefont {V.}~\bibnamefont
  {Voronyuk}}, \bibinfo {author} {\bibfnamefont {V.~D.}\ \bibnamefont
  {Toneev}}, \bibinfo {author} {\bibfnamefont {S.~A.}\ \bibnamefont
  {Voloshin}}, \ and\ \bibinfo {author} {\bibfnamefont {W.}~\bibnamefont
  {Cassing}},\ }\bibfield  {title} {\enquote {\bibinfo {title}
  {{Charge-dependent directed flow in asymmetric nuclear collisions}},}\
  }\href@noop {} {\bibfield  {journal} {\bibinfo  {journal} {Phys. Rev.}\
  }\textbf {\bibinfo {volume} {C90}},\ \bibinfo {pages} {064903} (\bibinfo
  {year} {2014})}\BibitemShut {NoStop}%
\bibitem [{\citenamefont {Deng}\ and\ \citenamefont {Huang}(2015)}]{deng}%
  \BibitemOpen
  \bibfield  {author} {\bibinfo {author} {\bibfnamefont {W.}~\bibnamefont
  {Deng}}\ and\ \bibinfo {author} {\bibfnamefont {X.}~\bibnamefont {Huang}},\
  }\bibfield  {title} {\enquote {\bibinfo {title} {{Electric fields and chiral
  magnetic effects in Cu+Au collisions}},}\ }\href@noop {} {\bibfield
  {journal} {\bibinfo  {journal} {Phys. Lett.}\ }\textbf {\bibinfo {volume}
  {B742}},\ \bibinfo {pages} {296--302} (\bibinfo {year} {2015})}\BibitemShut
  {NoStop}%
\bibitem [{\citenamefont {Pratt}(2013)}]{2wave1}%
  \BibitemOpen
  \bibfield  {author} {\bibinfo {author} {\bibfnamefont {S.}~\bibnamefont
  {Pratt}},\ }\bibfield  {title} {\enquote {\bibinfo {title} {{Viewing the
  Chemical Evolution of the Quark-Gluon Plasma with Charge Balance
  Functions}},}\ }\href@noop {} {\bibfield  {journal} {\bibinfo  {journal}
  {PoS}\ }\textbf {\bibinfo {volume} {CPOD2013}},\ \bibinfo {pages} {023}
  (\bibinfo {year} {2013})}\BibitemShut {NoStop}%
\bibitem [{\citenamefont {Bass}\ \emph {et~al.}(2000)\citenamefont {Bass},
  \citenamefont {Danielewicz},\ and\ \citenamefont {Pratt}}]{2wave2}%
  \BibitemOpen
  \bibfield  {author} {\bibinfo {author} {\bibfnamefont {S.~A.}\ \bibnamefont
  {Bass}}, \bibinfo {author} {\bibfnamefont {P.}~\bibnamefont {Danielewicz}}, \
  and\ \bibinfo {author} {\bibfnamefont {S.}~\bibnamefont {Pratt}},\ }\bibfield
   {title} {\enquote {\bibinfo {title} {{Clocking Hadronization in Relativistic
  Heavy-Ion Collisions with Balance Functions}},}\ }\href@noop {} {\bibfield
  {journal} {\bibinfo  {journal} {Phys. Rev. Lett.}\ }\textbf {\bibinfo
  {volume} {85}},\ \bibinfo {pages} {2689} (\bibinfo {year}
  {2000})}\BibitemShut {NoStop}%
\bibitem [{\citenamefont {Kharzeev}\ \emph {et~al.}(1998)\citenamefont
  {Kharzeev}, \citenamefont {Pisarski},\ and\ \citenamefont {Tytgat}}]{CME1}%
  \BibitemOpen
  \bibfield  {author} {\bibinfo {author} {\bibfnamefont {D.}~\bibnamefont
  {Kharzeev}}, \bibinfo {author} {\bibfnamefont {R.~D.}\ \bibnamefont
  {Pisarski}}, \ and\ \bibinfo {author} {\bibfnamefont {M.~H.~G.}\ \bibnamefont
  {Tytgat}},\ }\bibfield  {title} {\enquote {\bibinfo {title} {{Possibility of
  spontaneous parity violation in hot QCD}},}\ }\href@noop {} {\bibfield
  {journal} {\bibinfo  {journal} {Phys. Rev. Lett.}\ }\textbf {\bibinfo
  {volume} {81}},\ \bibinfo {pages} {512--515} (\bibinfo {year}
  {1998})}\BibitemShut {NoStop}%
\bibitem [{\citenamefont {Kharzeev}\ and\ \citenamefont
  {Pisarski}(2000)}]{CME2}%
  \BibitemOpen
  \bibfield  {author} {\bibinfo {author} {\bibfnamefont {D.}~\bibnamefont
  {Kharzeev}}\ and\ \bibinfo {author} {\bibfnamefont {R.~D.}\ \bibnamefont
  {Pisarski}},\ }\bibfield  {title} {\enquote {\bibinfo {title} {{Pionic
  measures of parity and CP violation in high-energy nuclear collisions}},}\
  }\href@noop {} {\bibfield  {journal} {\bibinfo  {journal} {Phys. Rev.}\
  }\textbf {\bibinfo {volume} {D61}},\ \bibinfo {pages} {111901} (\bibinfo
  {year} {2000})}\BibitemShut {NoStop}%
\bibitem [{\citenamefont {Kharzeev}\ and\ \citenamefont {Yee}(2011)}]{CMW_org}%
  \BibitemOpen
  \bibfield  {author} {\bibinfo {author} {\bibfnamefont {D.~E.}\ \bibnamefont
  {Kharzeev}}\ and\ \bibinfo {author} {\bibfnamefont {H.U.}\ \bibnamefont
  {Yee}},\ }\bibfield  {title} {\enquote {\bibinfo {title} {{Chiral Magnetic
  Wave}},}\ }\href@noop {} {\bibfield  {journal} {\bibinfo  {journal} {Phys.
  Rev.}\ }\textbf {\bibinfo {volume} {D83}},\ \bibinfo {pages} {085007}
  (\bibinfo {year} {2011})}\BibitemShut {NoStop}%
\bibitem [{\citenamefont {Burnier}\ \emph {et~al.}(2011)\citenamefont
  {Burnier}, \citenamefont {Kharzeev}, \citenamefont {Liao},\ and\
  \citenamefont {Yee}}]{CMW2}%
  \BibitemOpen
  \bibfield  {author} {\bibinfo {author} {\bibfnamefont {Y.}~\bibnamefont
  {Burnier}}, \bibinfo {author} {\bibfnamefont {D.~E.}\ \bibnamefont
  {Kharzeev}}, \bibinfo {author} {\bibfnamefont {J.}~\bibnamefont {Liao}}, \
  and\ \bibinfo {author} {\bibfnamefont {H.U.}\ \bibnamefont {Yee}},\
  }\bibfield  {title} {\enquote {\bibinfo {title} {{Chiral magnetic wave at
  finite baryon density and the electric quadrupole moment of quark-gluon
  plasma in heavy ion collisions}},}\ }\href@noop {} {\bibfield  {journal}
  {\bibinfo  {journal} {Phys. Rev. Lett.}\ }\textbf {\bibinfo {volume} {107}},\
  \bibinfo {pages} {052303} (\bibinfo {year} {2011})}\BibitemShut {NoStop}%
\bibitem [{\citenamefont {Kharzeev}\ \emph {et~al.}(2016)\citenamefont
  {Kharzeev}, \citenamefont {Liao}, \citenamefont {Voloshin},\ and\
  \citenamefont {Wang}}]{Kharzeev:2015znc}%
  \BibitemOpen
  \bibfield  {author} {\bibinfo {author} {\bibfnamefont {D.~E.}\ \bibnamefont
  {Kharzeev}}, \bibinfo {author} {\bibfnamefont {J.}~\bibnamefont {Liao}},
  \bibinfo {author} {\bibfnamefont {S.~A.}\ \bibnamefont {Voloshin}}, \ and\
  \bibinfo {author} {\bibfnamefont {G.}~\bibnamefont {Wang}},\ }\bibfield
  {title} {\enquote {\bibinfo {title} {{Chiral magnetic and vortical effects in
  high-energy nuclear collisions—A status report}},}\ }\href@noop {}
  {\bibfield  {journal} {\bibinfo  {journal} {Prog. Part. Nucl. Phys.}\
  }\textbf {\bibinfo {volume} {88}},\ \bibinfo {pages} {1} (\bibinfo {year}
  {2016})}\BibitemShut {NoStop}%
\bibitem [{\citenamefont {{B. I. Abelev {\it et al.} (STAR
  Collaboration)}}(2009)}]{CME_star}%
  \BibitemOpen
  \bibfield  {author} {\bibinfo {author} {\bibnamefont {{B. I. Abelev {\it et
  al.} (STAR Collaboration)}}},\ }\bibfield  {title} {\enquote {\bibinfo
  {title} {{Azimuthal Charged-Particle Correlations and Possible Local Strong
  Parity Violation}},}\ }\href@noop {} {\bibfield  {journal} {\bibinfo
  {journal} {Phys. Rev. Lett.}\ }\textbf {\bibinfo {volume} {103}},\ \bibinfo
  {pages} {251601} (\bibinfo {year} {2009})}\BibitemShut {NoStop}%
\bibitem [{\citenamefont {{B. Abelev {\it et al.} (ALICE
  Collaboration)}}(2013{\natexlab{a}})}]{CME_alice}%
  \BibitemOpen
  \bibfield  {author} {\bibinfo {author} {\bibnamefont {{B. Abelev {\it et al.}
  (ALICE Collaboration)}}},\ }\bibfield  {title} {\enquote {\bibinfo {title}
  {{Charge separation relative to the reaction plane in Pb-Pb collisions at
  $\sqrt{s_{NN}}= 2.76$ TeV}},}\ }\href@noop {} {\bibfield  {journal} {\bibinfo
   {journal} {Phys. Rev. Lett.}\ }\textbf {\bibinfo {volume} {110}},\ \bibinfo
  {pages} {012301} (\bibinfo {year} {2013}{\natexlab{a}})}\BibitemShut
  {NoStop}%
\bibitem [{\citenamefont {{L. Adamczyk {\it et al.} (STAR
  Collaboration)}}(2015)}]{CMW_star}%
  \BibitemOpen
  \bibfield  {author} {\bibinfo {author} {\bibnamefont {{L. Adamczyk {\it et
  al.} (STAR Collaboration)}}},\ }\bibfield  {title} {\enquote {\bibinfo
  {title} {{Observation of Charge Asymmetry Dependence of Pion Elliptic Flow
  and the Possible Chiral magnetic Wave in Heavy Ion Collisions}},}\
  }\href@noop {} {\bibfield  {journal} {\bibinfo  {journal} {Phys. Rev. Lett.}\
  }\textbf {\bibinfo {volume} {114}},\ \bibinfo {pages} {252302} (\bibinfo
  {year} {2015})}\BibitemShut {NoStop}%
\bibitem [{\citenamefont {{J. Adam {\it et al.} (ALICE
  Collaboration)}}(2016)}]{CMW_alice}%
  \BibitemOpen
  \bibfield  {author} {\bibinfo {author} {\bibnamefont {{J. Adam {\it et al.}
  (ALICE Collaboration)}}},\ }\bibfield  {title} {\enquote {\bibinfo {title}
  {{Charge-dependent flow and the search for the chiral magnetic wave in Pb-Pb
  collisions at $\sqrt{s_{NN}}$ = 2.76 TeV}},}\ }\href@noop {} {\bibfield
  {journal} {\bibinfo  {journal} {Phys. Rev.}\ }\textbf {\bibinfo {volume}
  {C93}},\ \bibinfo {pages} {044903} (\bibinfo {year} {2016})}\BibitemShut
  {NoStop}%
\bibitem [{\citenamefont {{M. Anderson {\it et al.}}}(2003)}]{tpc}%
  \BibitemOpen
  \bibfield  {author} {\bibinfo {author} {\bibnamefont {{M. Anderson {\it et
  al.}}}},\ }\bibfield  {title} {\enquote {\bibinfo {title} {{The Star time
  projection chamber: A unique tool for studying high multiplicity events at
  RHIC}},}\ }\href@noop {} {\bibfield  {journal} {\bibinfo  {journal} {Nucl.
  Instrum. Meth.}\ }\textbf {\bibinfo {volume} {A499}},\ \bibinfo {pages}
  {659--678} (\bibinfo {year} {2003})}\BibitemShut {NoStop}%
\bibitem [{\citenamefont {{W. J. Llope {\it et al.}}}(2014)}]{vpd}%
  \BibitemOpen
  \bibfield  {author} {\bibinfo {author} {\bibnamefont {{W. J. Llope {\it et
  al.}}}},\ }\bibfield  {title} {\enquote {\bibinfo {title} {{The STAR Vertex
  Position Detector}},}\ }\href@noop {} {\bibfield  {journal} {\bibinfo
  {journal} {Nucl. Instrum. Meth.}\ }\textbf {\bibinfo {volume} {A759}},\
  \bibinfo {pages} {23--28} (\bibinfo {year} {2014})}\BibitemShut {NoStop}%
\bibitem [{\citenamefont {Miller}\ \emph {et~al.}(2007)\citenamefont {Miller},
  \citenamefont {Reygers}, \citenamefont {Sanders},\ and\ \citenamefont
  {Steinberg}}]{glauber}%
  \BibitemOpen
  \bibfield  {author} {\bibinfo {author} {\bibfnamefont {M.~L.}\ \bibnamefont
  {Miller}}, \bibinfo {author} {\bibfnamefont {K.}~\bibnamefont {Reygers}},
  \bibinfo {author} {\bibfnamefont {S.~J.}\ \bibnamefont {Sanders}}, \ and\
  \bibinfo {author} {\bibfnamefont {P.}~\bibnamefont {Steinberg}},\ }\bibfield
  {title} {\enquote {\bibinfo {title} {{Glauber Modeling in High-Energy Nuclear
  Collisions}},}\ }\href@noop {} {\bibfield  {journal} {\bibinfo  {journal}
  {Ann. Rev. Nucl. Part. Sci.}\ }\textbf {\bibinfo {volume} {57}},\ \bibinfo
  {pages} {205} (\bibinfo {year} {2007})}\BibitemShut {NoStop}%
\bibitem [{\citenamefont {{L. Adamczyk {\it et al.} (STAR
  Collaboration)}}(2012)}]{BESv2}%
  \BibitemOpen
  \bibfield  {author} {\bibinfo {author} {\bibnamefont {{L. Adamczyk {\it et
  al.} (STAR Collaboration)}}},\ }\bibfield  {title} {\enquote {\bibinfo
  {title} {{Inclusive charged hadron elliptic flow in Au+Au collisions at
  $\sqrt{s_{_{NN}}}$ = 7.7 - 39 GeV}},}\ }\href@noop {} {\bibfield  {journal}
  {\bibinfo  {journal} {Phys. Rev.}\ }\textbf {\bibinfo {volume} {C86}},\
  \bibinfo {pages} {054908} (\bibinfo {year} {2012})}\BibitemShut {NoStop}%
\bibitem [{\citenamefont {{C. Adler {\it et al.}}}(2001)}]{zdc}%
  \BibitemOpen
  \bibfield  {author} {\bibinfo {author} {\bibnamefont {{C. Adler {\it et
  al.}}}},\ }\bibfield  {title} {\enquote {\bibinfo {title} {{The RHIC zero
  degree calorimeters}},}\ }\href@noop {} {\bibfield  {journal} {\bibinfo
  {journal} {Nucl. Instrum. Meth.}\ }\textbf {\bibinfo {volume} {A461}},\
  \bibinfo {pages} {337--340} (\bibinfo {year} {2001})}\BibitemShut {NoStop}%
\bibitem [{\citenamefont {{J. Adams {\it et al.} (STAR
  Collaboration)}}(2006)}]{v1smd}%
  \BibitemOpen
  \bibfield  {author} {\bibinfo {author} {\bibnamefont {{J. Adams {\it et al.}
  (STAR Collaboration)}}},\ }\bibfield  {title} {\enquote {\bibinfo {title}
  {{Directed flow in Au+Au collisions at $\sqrt{s_{NN}}$ = 62.4 GeV}},}\
  }\href@noop {} {\bibfield  {journal} {\bibinfo  {journal} {Phys. Rev.}\
  }\textbf {\bibinfo {volume} {C73}},\ \bibinfo {pages} {034903} (\bibinfo
  {year} {2006})}\BibitemShut {NoStop}%
\bibitem [{\citenamefont {{B. I. Abelev {\it et al.} (STAR
  Collaboration)}}(2008)}]{v1smd_star}%
  \BibitemOpen
  \bibfield  {author} {\bibinfo {author} {\bibnamefont {{B. I. Abelev {\it et
  al.} (STAR Collaboration)}}},\ }\bibfield  {title} {\enquote {\bibinfo
  {title} {{System-Size Independence of Directed Flow Measured at the BNL
  Relativistic Heavy Ion Collider}},}\ }\href@noop {} {\bibfield  {journal}
  {\bibinfo  {journal} {Phys. Rev. Lett.}\ }\textbf {\bibinfo {volume} {101}},\
  \bibinfo {pages} {252301} (\bibinfo {year} {2008})}\BibitemShut {NoStop}%
\bibitem [{\citenamefont {Voloshin}\ and\ \citenamefont
  {Niida}(2016)}]{spflow}%
  \BibitemOpen
  \bibfield  {author} {\bibinfo {author} {\bibfnamefont {S.~A.}\ \bibnamefont
  {Voloshin}}\ and\ \bibinfo {author} {\bibfnamefont {T.}~\bibnamefont
  {Niida}},\ }\bibfield  {title} {\enquote {\bibinfo {title}
  {{Ultra-relativistic nuclear collisions: where the spectator flow?}}}\
  }\href@noop {} {\bibfield  {journal} {\bibinfo  {journal} {arXiv:1604.04597}\
  } (\bibinfo {year} {2016})}\BibitemShut {NoStop}%
\bibitem [{\citenamefont {Poskanzer}\ and\ \citenamefont
  {Voloshin}(1998)}]{TwoSub}%
  \BibitemOpen
  \bibfield  {author} {\bibinfo {author} {\bibfnamefont {A.~M.}\ \bibnamefont
  {Poskanzer}}\ and\ \bibinfo {author} {\bibfnamefont {S.~A.}\ \bibnamefont
  {Voloshin}},\ }\bibfield  {title} {\enquote {\bibinfo {title} {{Methods for
  analyzing anisotropic flow in relativistic nuclear collisions}},}\
  }\href@noop {} {\bibfield  {journal} {\bibinfo  {journal} {Phys. Rev.}\
  }\textbf {\bibinfo {volume} {C58}},\ \bibinfo {pages} {1671} (\bibinfo {year}
  {1998})}\BibitemShut {NoStop}%
\bibitem [{\citenamefont {Teaney}\ and\ \citenamefont {Yan}(2011)}]{Teaney}%
  \BibitemOpen
  \bibfield  {author} {\bibinfo {author} {\bibfnamefont {D.}~\bibnamefont
  {Teaney}}\ and\ \bibinfo {author} {\bibfnamefont {L.}~\bibnamefont {Yan}},\
  }\bibfield  {title} {\enquote {\bibinfo {title} {{Triangularity and dipole
  asymmetry in relativistic heavy ion collisions}},}\ }\href@noop {} {\bibfield
   {journal} {\bibinfo  {journal} {Phys. Rev.}\ }\textbf {\bibinfo {volume}
  {C83}},\ \bibinfo {pages} {064904} (\bibinfo {year} {2011})}\BibitemShut
  {NoStop}%
\bibitem [{\citenamefont {Luzum}\ and\ \citenamefont
  {Ollitrault}(2011)}]{Luzum}%
  \BibitemOpen
  \bibfield  {author} {\bibinfo {author} {\bibfnamefont {M.}~\bibnamefont
  {Luzum}}\ and\ \bibinfo {author} {\bibfnamefont {J.Y.}\ \bibnamefont
  {Ollitrault}},\ }\bibfield  {title} {\enquote {\bibinfo {title} {{Directed
  Flow at Midrapidity in Heavy-Ion Collisions}},}\ }\href@noop {} {\bibfield
  {journal} {\bibinfo  {journal} {Phys. Rev. Lett.}\ }\textbf {\bibinfo
  {volume} {106}},\ \bibinfo {pages} {102301} (\bibinfo {year}
  {2011})}\BibitemShut {NoStop}%
\bibitem [{\citenamefont {{B. Abelev {\it et al.} (ALICE
  Collaboration)}}(2013{\natexlab{b}})}]{aliceV1}%
  \BibitemOpen
  \bibfield  {author} {\bibinfo {author} {\bibnamefont {{B. Abelev {\it et al.}
  (ALICE Collaboration)}}},\ }\bibfield  {title} {\enquote {\bibinfo {title}
  {{Directed flow of charged particles at midrapidity relative to the spectator
  plane in Pb+Pb collisions at $\sqrt{s_{_{NN}}}$ = 2.76 TeV}},}\ }\href@noop
  {} {\bibfield  {journal} {\bibinfo  {journal} {Phys. Rev. Lett.}\ }\textbf
  {\bibinfo {volume} {111}},\ \bibinfo {pages} {232302} (\bibinfo {year}
  {2013}{\natexlab{b}})}\BibitemShut {NoStop}%
\bibitem [{\citenamefont {{C. A. Whitten for the STAR
  Collaboration}}(2008)}]{bbc}%
  \BibitemOpen
  \bibfield  {author} {\bibinfo {author} {\bibnamefont {{C. A. Whitten for the
  STAR Collaboration}}},\ }\bibfield  {title} {\enquote {\bibinfo {title} {{The
  beam-beam counter: A local polarimeter st STAR}},}\ }\href@noop {} {\bibfield
   {journal} {\bibinfo  {journal} {AIP Conf. Proc.}\ }\textbf {\bibinfo
  {volume} {980}},\ \bibinfo {pages} {390--396} (\bibinfo {year}
  {2008})}\BibitemShut {NoStop}%
\bibitem [{\citenamefont {{C. E. Allgower {\it et al.} (STAR
  Collaboration)}}(2003)}]{eemc}%
  \BibitemOpen
  \bibfield  {author} {\bibinfo {author} {\bibnamefont {{C. E. Allgower {\it et
  al.} (STAR Collaboration)}}},\ }\bibfield  {title} {\enquote {\bibinfo
  {title} {{The STAR endcap electromagnetic calorimeter}},}\ }\href@noop {}
  {\bibfield  {journal} {\bibinfo  {journal} {Nucl. Instrum. Meth.}\ }\textbf
  {\bibinfo {volume} {A499}},\ \bibinfo {pages} {740--750} (\bibinfo {year}
  {2003})}\BibitemShut {NoStop}%
\bibitem [{\citenamefont {Bo$\dot{z}$ek}(2012)}]{bozek}%
  \BibitemOpen
  \bibfield  {author} {\bibinfo {author} {\bibfnamefont {P.}~\bibnamefont
  {Bo$\dot{z}$ek}},\ }\bibfield  {title} {\enquote {\bibinfo {title}
  {{Event-by-event viscous hydrodynamics for Cu-Au collisions at
  $\sqrt{s_{_{NN}}}$ = 200 GeV}},}\ }\href@noop {} {\bibfield  {journal}
  {\bibinfo  {journal} {Phys. Lett.}\ }\textbf {\bibinfo {volume} {B717}},\
  \bibinfo {pages} {287--290} (\bibinfo {year} {2012})}\BibitemShut {NoStop}%
\bibitem [{\citenamefont {Heinz}\ and\ \citenamefont {Kolb}(2004)}]{v1_heinz}%
  \BibitemOpen
  \bibfield  {author} {\bibinfo {author} {\bibfnamefont {U.~W.}\ \bibnamefont
  {Heinz}}\ and\ \bibinfo {author} {\bibfnamefont {P.~F.}\ \bibnamefont
  {Kolb}},\ }\bibfield  {title} {\enquote {\bibinfo {title} {{Rapidity
  dependent momentum anisotropy at RHIC}},}\ }\href@noop {} {\bibfield
  {journal} {\bibinfo  {journal} {J. Phys.}\ }\textbf {\bibinfo {volume}
  {G30}},\ \bibinfo {pages} {S1229--S1234} (\bibinfo {year}
  {2004})}\BibitemShut {NoStop}%
\bibitem [{\citenamefont {{A. Adare {\it et al.} (PHENIX
  Collaboration)}}(2015)}]{cuauphenix}%
  \BibitemOpen
  \bibfield  {author} {\bibinfo {author} {\bibnamefont {{A. Adare {\it et al.}
  (PHENIX Collaboration)}}},\ }\bibfield  {title} {\enquote {\bibinfo {title}
  {{Measurements of directed, elliptic, and triangular flow in Cu+Au collisions
  at $\sqrt{s_{_{NN}}}$ = 200 GeV}},}\ }\href@noop {} {\bibfield  {journal}
  {\bibinfo  {journal} {arXiv:1509.07784}\ } (\bibinfo {year}
  {2015})}\BibitemShut {NoStop}%
\bibitem [{\citenamefont {Gupta}(2004)}]{Gupta}%
  \BibitemOpen
  \bibfield  {author} {\bibinfo {author} {\bibfnamefont {S.}~\bibnamefont
  {Gupta}},\ }\bibfield  {title} {\enquote {\bibinfo {title} {{The Electrical
  conductivity and soft photon emissivity of the QCD plasma}},}\ }\href@noop {}
  {\bibfield  {journal} {\bibinfo  {journal} {Phys. Lett.}\ }\textbf {\bibinfo
  {volume} {B597}},\ \bibinfo {pages} {57--62} (\bibinfo {year}
  {2004})}\BibitemShut {NoStop}%
\bibitem [{\citenamefont {Aarts}\ \emph {et~al.}(2007)\citenamefont {Aarts},
  \citenamefont {Allton}, \citenamefont {Foley}, \citenamefont {Hands},\ and\
  \citenamefont {Kim}}]{Aarts}%
  \BibitemOpen
  \bibfield  {author} {\bibinfo {author} {\bibfnamefont {G.}~\bibnamefont
  {Aarts}}, \bibinfo {author} {\bibfnamefont {C.}~\bibnamefont {Allton}},
  \bibinfo {author} {\bibfnamefont {J.}~\bibnamefont {Foley}}, \bibinfo
  {author} {\bibfnamefont {S.}~\bibnamefont {Hands}}, \ and\ \bibinfo {author}
  {\bibfnamefont {S.}~\bibnamefont {Kim}},\ }\bibfield  {title} {\enquote
  {\bibinfo {title} {{Spectral Functions at Small Energies and the Electrical
  Conductivity in Hot Quenched Lattice QCD}},}\ }\href@noop {} {\bibfield
  {journal} {\bibinfo  {journal} {Phys. Rev. Lett.}\ }\textbf {\bibinfo
  {volume} {99}},\ \bibinfo {pages} {022002} (\bibinfo {year}
  {2007})}\BibitemShut {NoStop}%
\bibitem [{\citenamefont {Arnold}\ \emph {et~al.}(2003)\citenamefont {Arnold},
  \citenamefont {Moore},\ and\ \citenamefont {Yaffe}}]{pQCD}%
  \BibitemOpen
  \bibfield  {author} {\bibinfo {author} {\bibfnamefont {P.}~\bibnamefont
  {Arnold}}, \bibinfo {author} {\bibfnamefont {G.~D.}\ \bibnamefont {Moore}}, \
  and\ \bibinfo {author} {\bibfnamefont {L.~G.}\ \bibnamefont {Yaffe}},\
  }\bibfield  {title} {\enquote {\bibinfo {title} {{Transport coefficients in
  high temperature gauge theories, 2. Beyond leading log}},}\ }\href@noop {}
  {\bibfield  {journal} {\bibinfo  {journal} {JHEP}\ }\textbf {\bibinfo
  {volume} {05}},\ \bibinfo {pages} {051} (\bibinfo {year} {2003})}\BibitemShut
  {NoStop}%
\bibitem [{\citenamefont {Greif}\ \emph {et~al.}(2014)\citenamefont {Greif},
  \citenamefont {Bouras}, \citenamefont {Greiner},\ and\ \citenamefont
  {Xu}}]{pQCD2}%
  \BibitemOpen
  \bibfield  {author} {\bibinfo {author} {\bibfnamefont {M.}~\bibnamefont
  {Greif}}, \bibinfo {author} {\bibfnamefont {I.}~\bibnamefont {Bouras}},
  \bibinfo {author} {\bibfnamefont {C.}~\bibnamefont {Greiner}}, \ and\
  \bibinfo {author} {\bibfnamefont {Z.}~\bibnamefont {Xu}},\ }\bibfield
  {title} {\enquote {\bibinfo {title} {{Electric conductivity of the
  quark-gluon plasma investigated using a perturbative QCD based parton
  cascade}},}\ }\href@noop {} {\bibfield  {journal} {\bibinfo  {journal} {Phys.
  Rev.}\ }\textbf {\bibinfo {volume} {D90}},\ \bibinfo {pages} {094014}
  (\bibinfo {year} {2014})}\BibitemShut {NoStop}%
\bibitem [{\citenamefont {Steinert}\ and\ \citenamefont
  {Cassing}(2014)}]{phsd_Sgm}%
  \BibitemOpen
  \bibfield  {author} {\bibinfo {author} {\bibfnamefont {T.}~\bibnamefont
  {Steinert}}\ and\ \bibinfo {author} {\bibfnamefont {W.}~\bibnamefont
  {Cassing}},\ }\bibfield  {title} {\enquote {\bibinfo {title} {{Electric and
  magnetic response of hot QCD matter}},}\ }\href@noop {} {\bibfield  {journal}
  {\bibinfo  {journal} {Phys. Rev.}\ }\textbf {\bibinfo {volume} {C89}},\
  \bibinfo {pages} {035203} (\bibinfo {year} {2014})}\BibitemShut {NoStop}%
\bibitem [{\citenamefont {{L. McLerrana and V. Skokov}}(2014)}]{McLerrana2014}%
  \BibitemOpen
  \bibfield  {author} {\bibinfo {author} {\bibnamefont {{L. McLerrana and V.
  Skokov}}},\ }\bibfield  {title} {\enquote {\bibinfo {title} {{Comments about
  the electromagnetic field in heavy-ion collisions}},}\ }\href@noop {}
  {\bibfield  {journal} {\bibinfo  {journal} {Nucl. Phys.}\ }\textbf {\bibinfo
  {volume} {A929}},\ \bibinfo {pages} {184–190} (\bibinfo {year}
  {2014})}\BibitemShut {NoStop}%
\bibitem [{\citenamefont {{K. Tuchin}}(2013)}]{KTuchin2013}%
  \BibitemOpen
  \bibfield  {author} {\bibinfo {author} {\bibnamefont {{K. Tuchin}}},\
  }\bibfield  {title} {\enquote {\bibinfo {title} {{Time and space dependence
  of the electromagnetic field in relativistic heavy-ion collisions}},}\
  }\href@noop {} {\bibfield  {journal} {\bibinfo  {journal} {Phys. Rev.}\
  }\textbf {\bibinfo {volume} {C88}},\ \bibinfo {pages} {024911} (\bibinfo
  {year} {2013})}\BibitemShut {NoStop}%
\bibitem [{\citenamefont {{B. G. Zakharov}}(2014)}]{BZakharov2014}%
  \BibitemOpen
  \bibfield  {author} {\bibinfo {author} {\bibnamefont {{B. G. Zakharov}}},\
  }\bibfield  {title} {\enquote {\bibinfo {title} {{Electromagnetic response of
  quark–gluon plasma in heavy-ion collisions}},}\ }\href@noop {} {\bibfield
  {journal} {\bibinfo  {journal} {Phys. Lett.}\ }\textbf {\bibinfo {volume}
  {B737}},\ \bibinfo {pages} {262–266} (\bibinfo {year} {2014})}\BibitemShut
  {NoStop}%
\bibitem [{\citenamefont {{HERAPDF1.5}}()}]{herapdf}%
  \BibitemOpen
  \bibfield  {author} {\bibinfo {author} {\bibnamefont {{HERAPDF1.5}}},\
  }\href@noop {} {}\bibinfo {howpublished}
  {\url{https://www.desy.de/h1zeus/combined_results/herapdftable/}}\BibitemShut
  {NoStop}%
\end{thebibliography}%
